      [1999/12/01 v1.4c Il Nuovo Cimento]
\documentclass{cimento}

\usepackage{graphicx}  
\usepackage[utf8x]{inputenc}
\usepackage{subfigure}
\usepackage{wrapfig}
\usepackage{sidecap}

\title{A New Strategy to Discover Heavy Colored Vectors at the Early LHC}
\author{Natascia Vignaroli\from{ins:x}}
\instlist{\inst{ins:x} Dipartimento di Fisica, Università di Roma ``La Sapienza'' and INFN, Sezione di Roma, P.le A. Moro 2, I-00185 Roma, Italy
}
\begin{document}

\maketitle

\begin{abstract}
We perform a study of the LHC discovery reach on a heavy gluon ($G^*$) and heavy fermions (top and bottom excitations), 
coming from a new composite sector. We find that heavy fermion resonances have a great impact on the composite gluon phenomenology. 
If the composite gluon is heavier than composite fermions, as flavor observables seem to suggest, 
the search in the channel where $G^*$ decays into a heavy fermion plus its Standard Model partner is very promising, 
with the possibility for both the $G^*$ and heavy fermions to be discovered at the early stage of the LHC. 
The channel offers also the possibility to extract important information on model parameters, such as the top degree of compositeness.  
\end{abstract}
\begin{figure}[t]
\mbox{\subfigure{\includegraphics[width=0.25\textwidth, angle=-90]{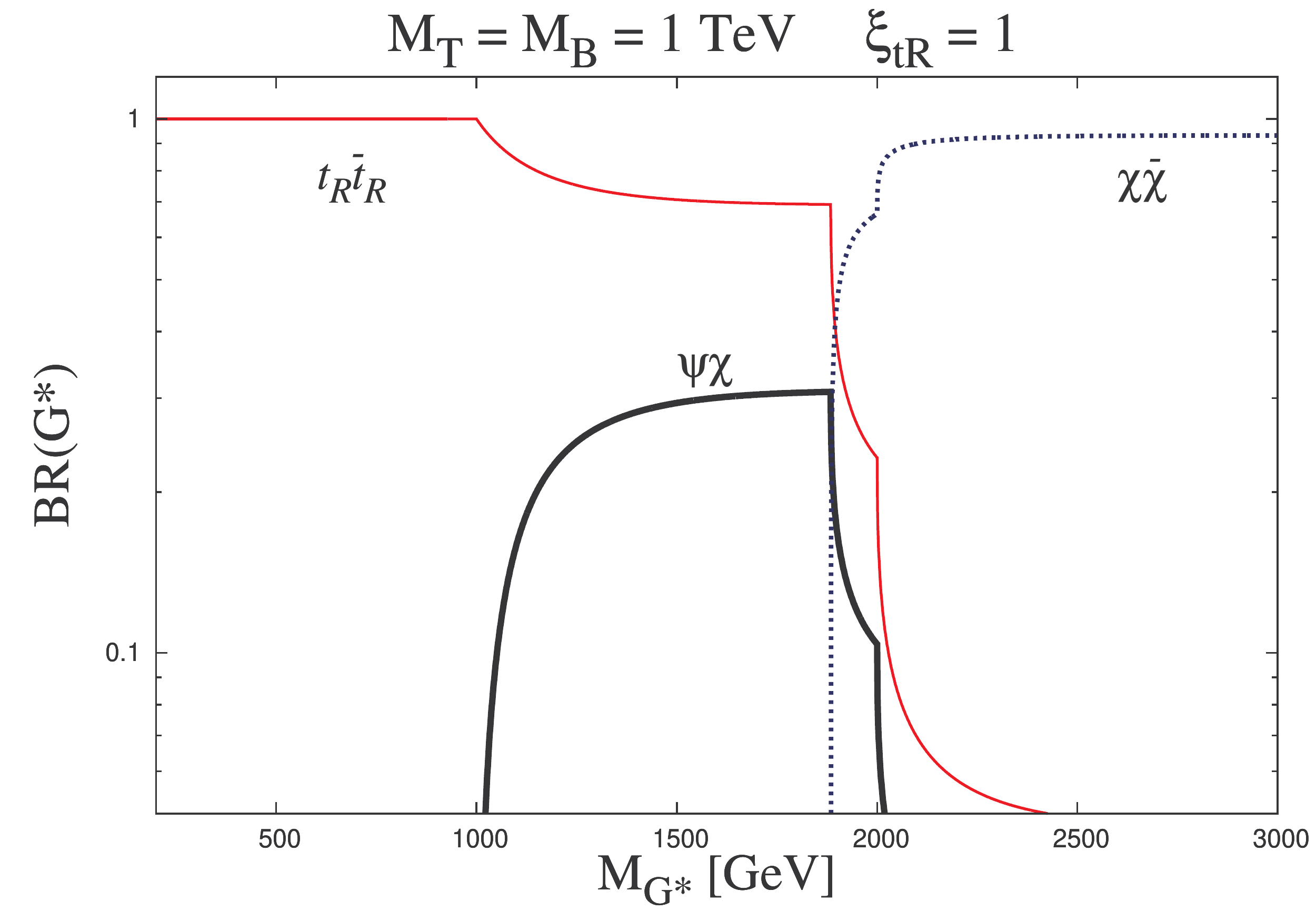}}\subfigure{\includegraphics[width=0.25\textwidth, angle=-90]{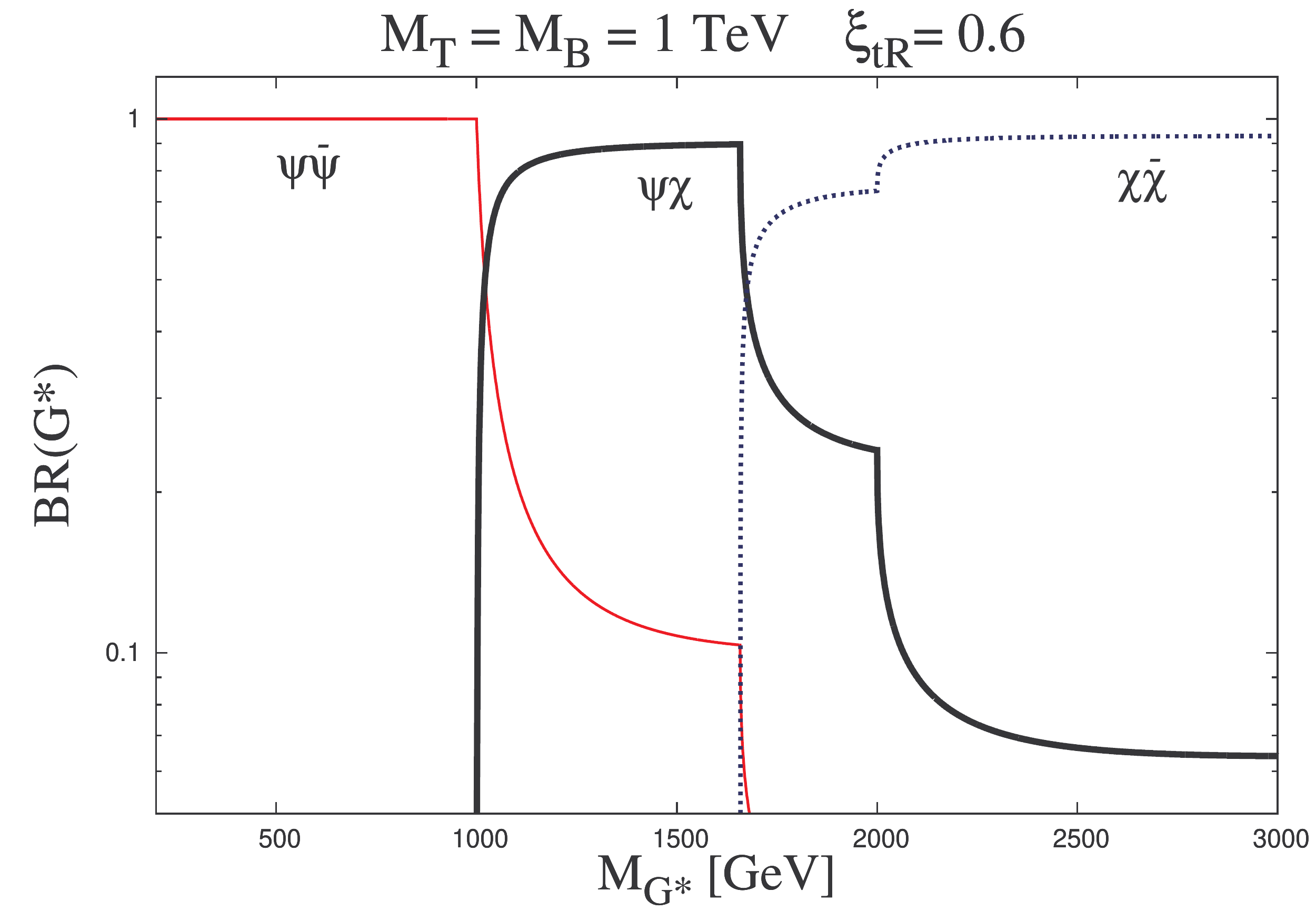}}
\subfigure{\includegraphics[width=0.25\textwidth, angle=-90]{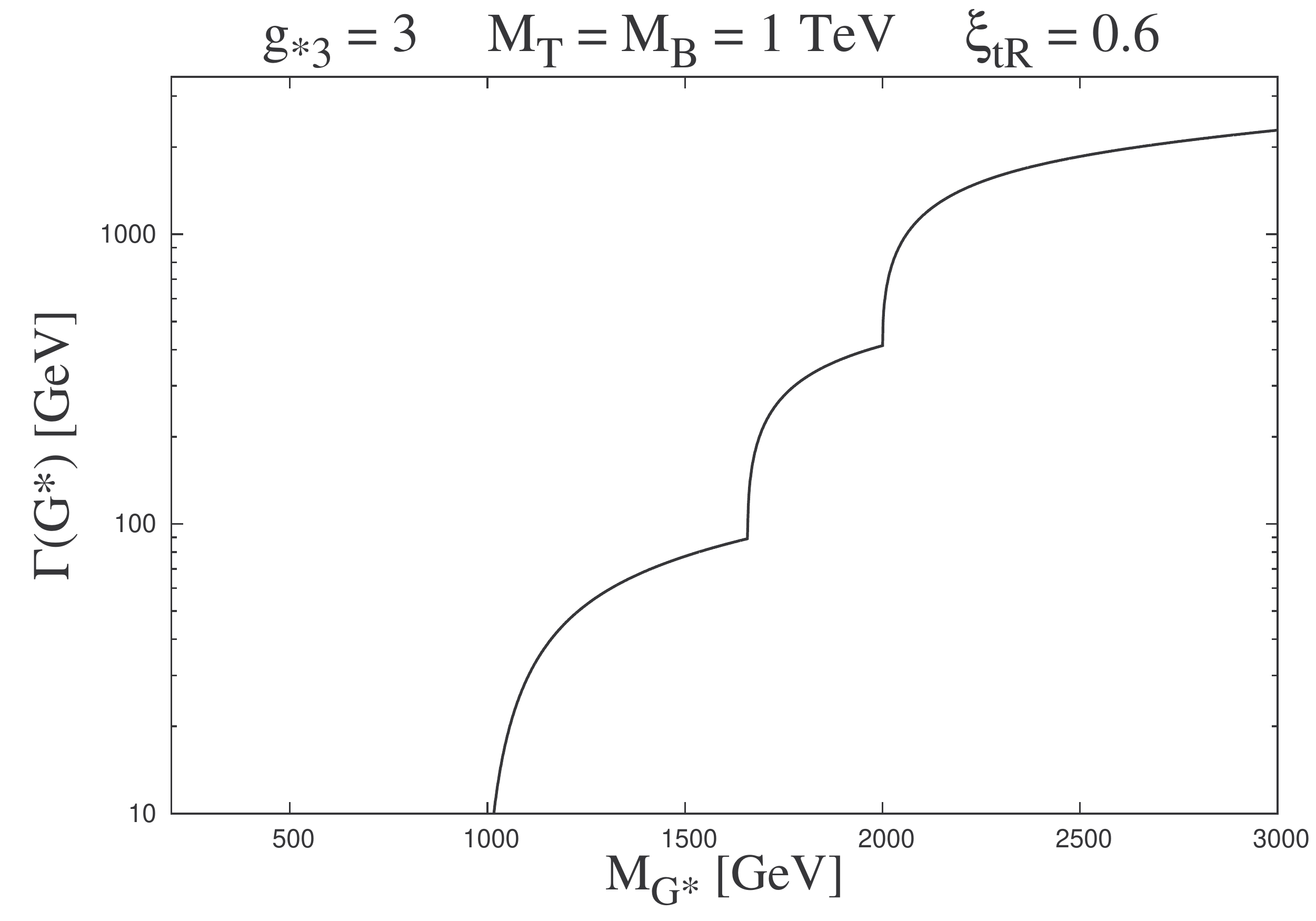}}}
\caption{$G^{*}$ decay BR for a fully composite top right, $\xi_{tR}=1$, and for an intermediate top degree of compositeness, $\xi_{tR}=0.6$, 
and $G^*$ total decay width (for $\xi_{tR}=0.6$), as functions of the $G^{*}$ mass.
The $(T,B)$ heavy fermions (partners of $q_L\equiv(t_L,b_L)$) mass has been set to $M_T=M_B=1$ TeV. 
$\psi\bar{\psi}$ denotes the BR for the $G^{*}$ decays into SM fermion pairs [red curve], 
$\psi\chi$ those for the $G^{*}$ decays into one heavy ($\chi$) plus one SM ($\psi$) fermion [thick curve] and $\chi\bar{\chi}$ 
those for the $G^{*}$ decays into a pair of heavy fermions [dotted curve]. 
}
\label{BR}
\end{figure}
\begin{figure}[h!]%
\begin{center}
\parbox{1.7in}{\includegraphics[width=0.31\textwidth]{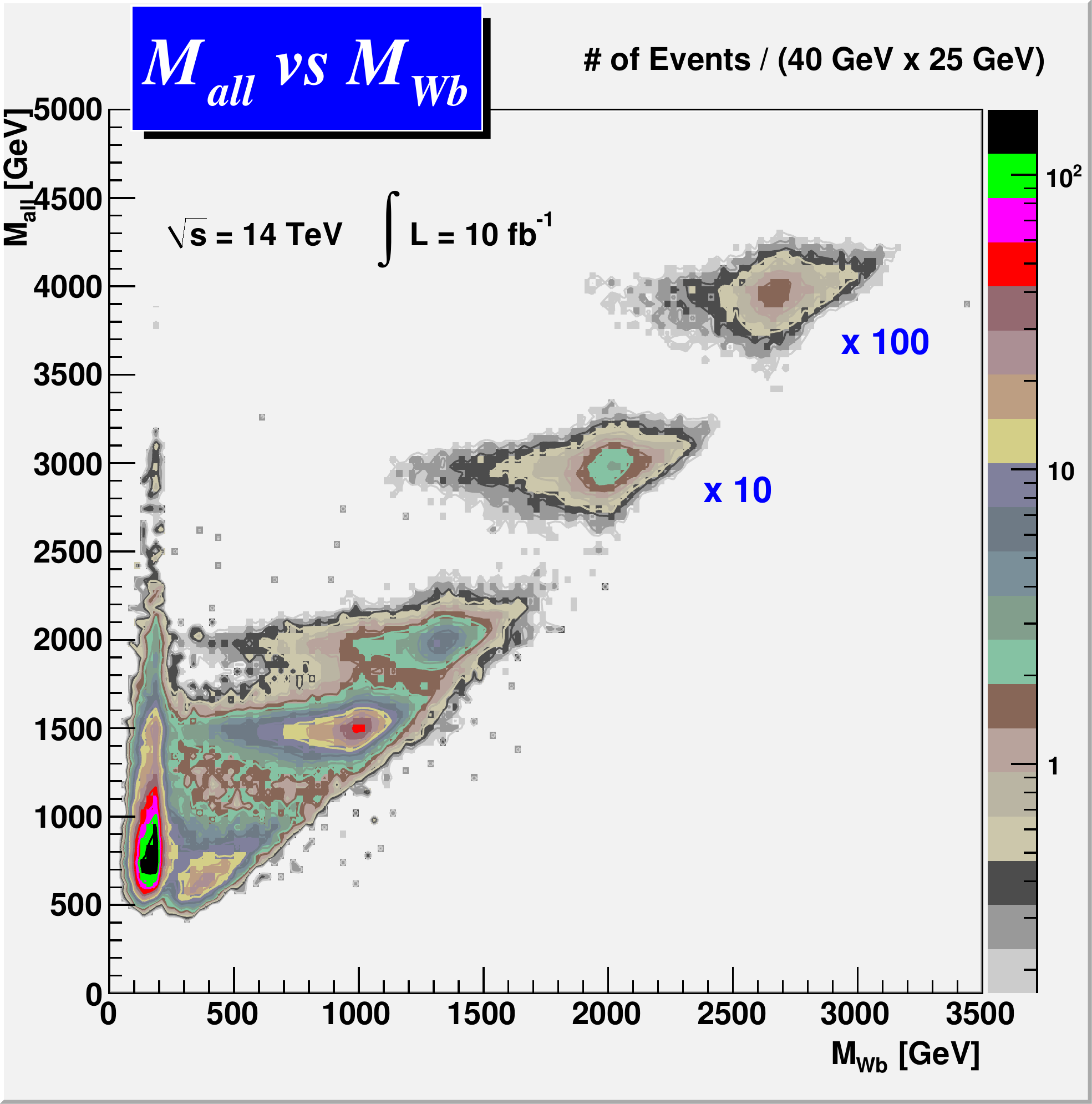}}%
\parbox{1.7in}{\includegraphics[width=0.31\textwidth]{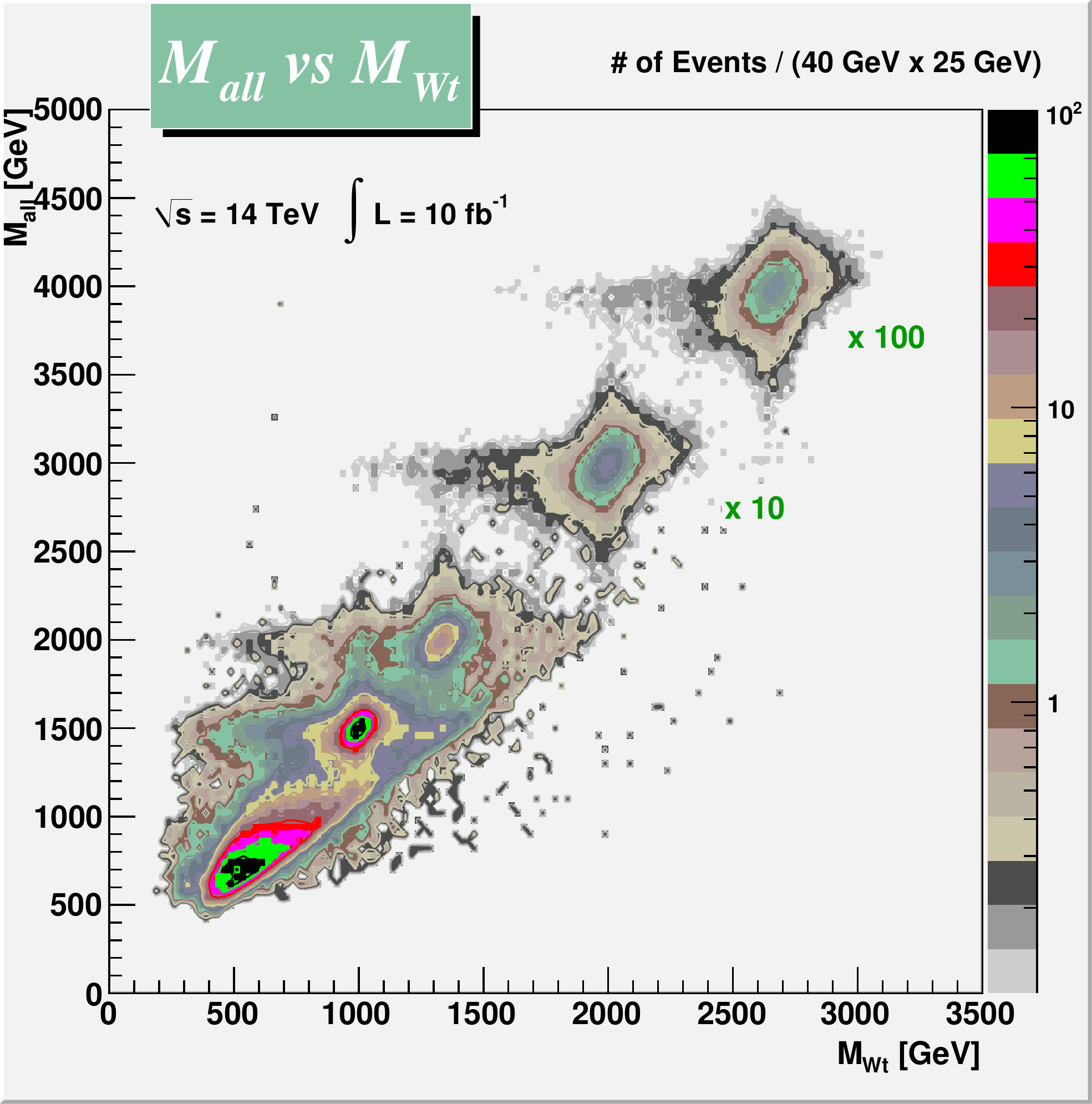}}%
\parbox{1.7in}{\includegraphics[width=0.31\textwidth]{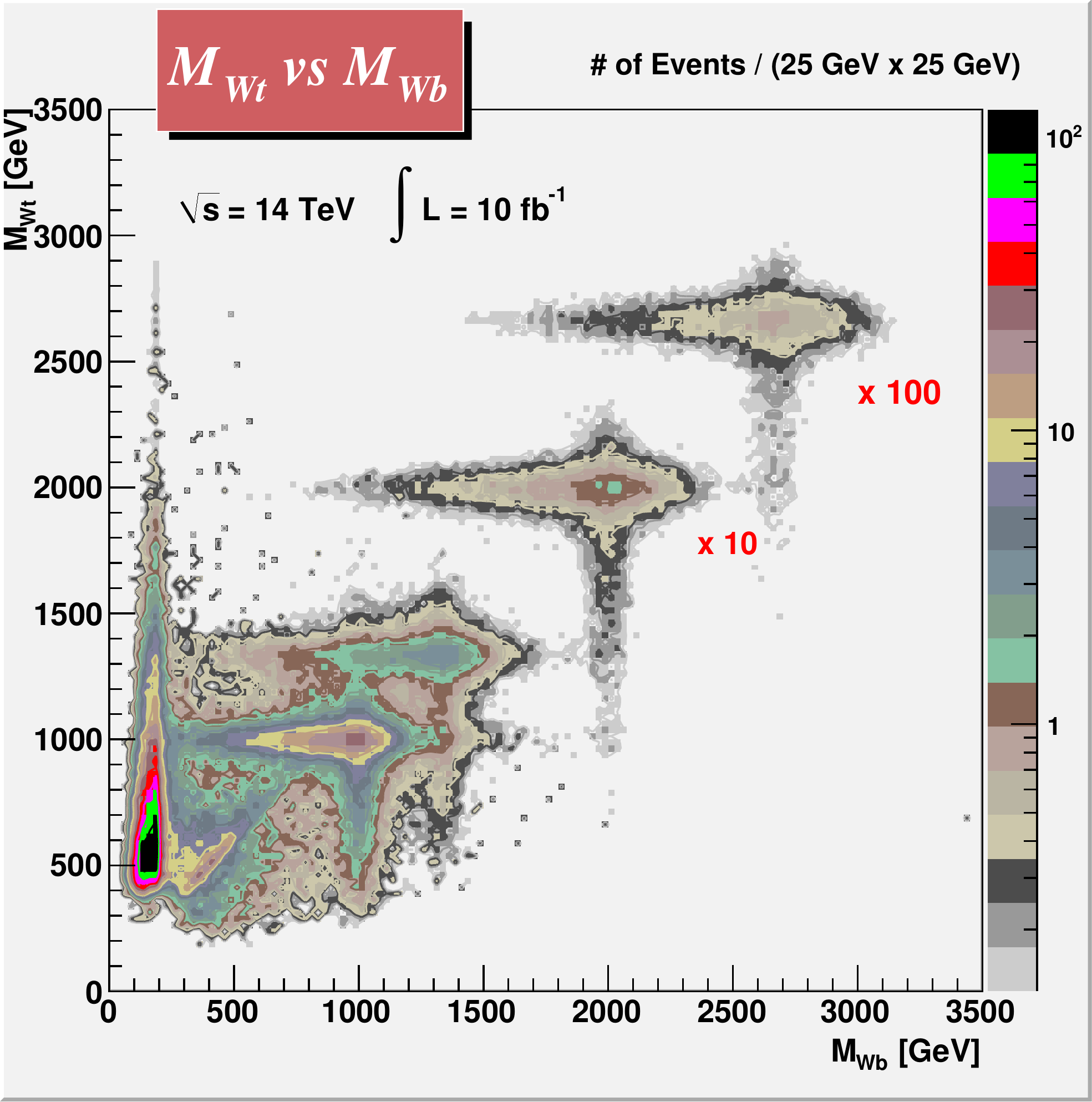}}%
\end{center}
\caption{Scatter Plots of invariant mass distributions,
 $M_{all}$ vs $M_{Wb}$, $M_{all}$ vs $M_{Wt}$ and $M_{Wt}$ vs $M_{Wb}$, 
for the Background and the Signal with $M_{G^*}=1.5,2,3,4$ TeV 
(and for $\xi_{tR}$=0.6 and a reference value of the composite strong coupling, $g_{*3}=3$).
}
\label{inv_mass}
\end{figure}

This analysis has been performed taking into account composite Higgs Models in a ``two-site'' (TS) description \cite{Sundrum}.
The heavy partner of the gluon has a large degree of compositeness and, as a consequence, it has
larger couplings to the heavier particles (which are also those with larger degrees of compositeness).
In the scenario where the $G^{∗}$ is below the threshold for the production of a heavy
fermion, $G^∗$ decays almost completely to top pairs. Until now, this first scenario is
the only one 
considered for the $G^∗$ search at the LHC \cite{Agashe}, but it seems to be not the preferred one by the data,
 that give generally stronger constraints on the $G^*$ mass than on the heavy fermion masses. If $G^*$ is heavier than fermionic resonances, the Branching Ratios (BR)
for the $G^*$ decays into a heavy fermion ($\chi$) plus its Standard Model partner ($\psi$) become important, as Fig. \ref{BR} shows, and they also increase in the case of a not fully composite right-handed top.
The analysis we will perform considering these decay channels is very promising because the presence of heavy fermion resonances in the Signal allows for a clean distinction
from the Background. We point out that there is also a pessimistic scenario, corresponding to the case of a very heavy $G^*$, with a mass 
greater than heavy fermion pairs. In this case the $G^*$ total decay width becomes too large (O(TeV)) to distinguish its resonance from the Background (Fig. \ref{BR}).\\
We analyze the $G^* \to \psi\chi$ decay channels. Heavy fermions decay into longitudinally polarized bosons (or into the Higgs)
 and we can identify three interesting search channels 
with the following final states: $Z(/h) t\bar{t}$, $Z(/h) b\bar{b}$ and $Wtb$. 
We will focus on the last, that has a component from the $G^*$ decays into a bottom plus its excited states ($B$, $\tilde{B}$),
\begin{wrapfigure}{l}{0.39\textwidth}
  \includegraphics[width=0.38\textwidth]{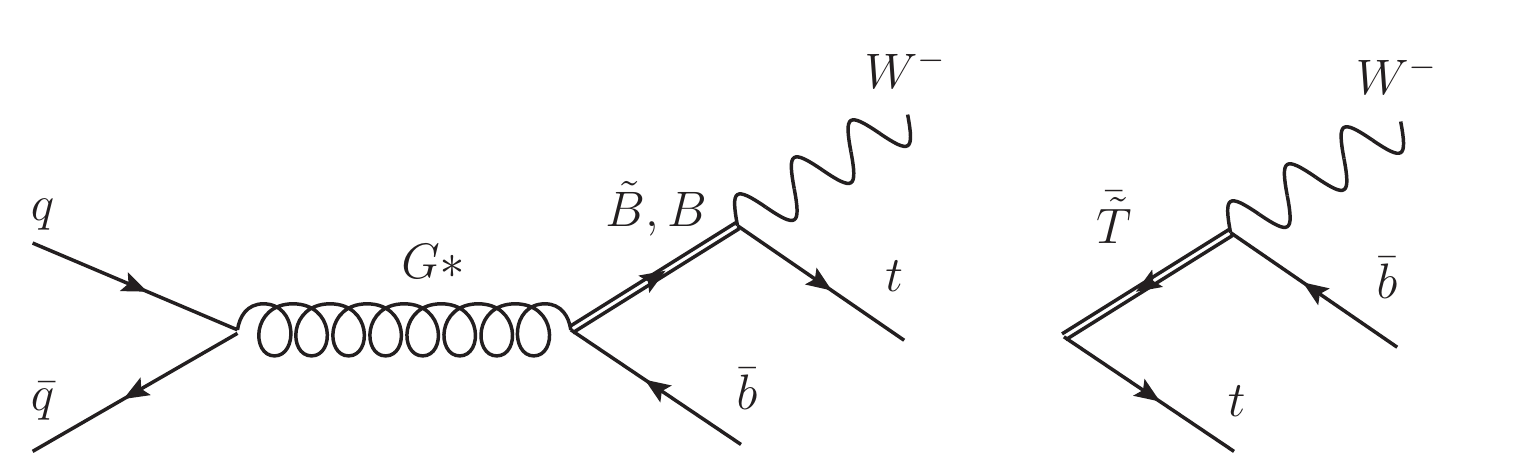}
\caption{$G^* \to \psi\chi$ Signal}
\end{wrapfigure}
from the top (right-handed) and its heavy partner ($\tilde{T}$)
and also 
from the `ordinary' $G^* \to t\bar{t}$ decay. We fix the ratio between the $G^*$ and the heavy fermions mass, 
$\frac{M_{G^*}}{m_*}=1.5$, and we look for a $G^*$ (and heavy fermions) Signal in the
semileptonic channel, $W(\to l\nu)W(\to jj)b\bar{b}$ ($l\equiv e/\mu$). The Background to our Signal comes mostly from $WWb\bar{b}$, 
other relevant backgrounds are $W(\to l\nu)b\bar{b}+Jets$ and $W(\to l\nu)+Jets$. We simulate events
 considering both $\sqrt{s} = 14$ TeV and $\sqrt{s} = 7$ TeV at the LHC.
After applying acceptance cuts\footnote{
At least 3 Jets (2 b-Jet) and 1 lepton obeying
$\Delta R_{jj} > 0.4 , \ \Delta R_{lj} > 0.4 , \  |\eta_j|<5 \ (|\eta_b|<2.5\ for\ the\ b-TAG), \ p_{T j}> 30$ GeV  ,  $|\eta_l|<2.5 , \  p_{T l}> 20$ GeV .
}, we reconstruct the neutrino\footnote{We obtain the neutrino $p_T$ from the missing transverse momentum and we require 
that the lepton and neutrino reconstruct an on-mass-shell W, $M_{l\nu}=80.4$ GeV.
This procedure gives us two values for the neutrino $p_z$; in the cases where neutrino comes from the decay of a top,
 we select the solution that gives the $M_{l\nu b}$ value closest to $174$ GeV.}, the $W$s 
and the one top\footnote{We reconstruct the leptonically and hadronically decayed $W$s and, by finding the $Wb$ pair
with the invariant mass closest to the top mass, the top.}. 
This allows us to calculate, besides the total invariant mass distribution, $M_{all}$, peaked, for the Signal, around the $G^*$ mass,
the invariant mass distributions, $M_{Wb}$ and $M_{Wt}$ (where the $W$ and the $b$ are not part of the reconstructed top), which are peaked, for the Signal, around
 the mass of the heavy fermions $\tilde{T}$ and $B / \tilde{B}$ respectively. 
We find particularly useful to look at scatter plots of invariant mass distributions,
 $M_{all}$ vs $M_{Wb}$, $M_{all}$ vs $M_{Wt}$ and $M_{Wb}$ vs $M_{Wt}$. This allows for a clean distinction between the Signal and the Background, 
which is predominantly distributed on small invariant mass values,
as Fig. \ref{inv_mass} shows\footnote{We calculate the invariant mass distributions after we applied several 'conservative' cuts. These are cuts in $p_T$ 
that reject less than the $3\%$ of the Signal, which is characterized by very energetic final states. We impose for the 14 (7) TeV analysis: 
$p_{Tj (1)} > 175 (155)$  GeV , $\ p_{Tj (2)} > 85 (75)$ GeV ,  $p_{Ttop} > 110 (105)$  GeV  ,  $p_{TW} > 110 (90)$ GeV ,  $p_{Tb} > 70 (65)$ GeV . 
 $j (1)$ denotes the jet (light-jet or b-jet) with the highest $p_T$ ($j (2)$ is the second most energetic jet); 
$W$ and $b$ do not come from the reconstructed top decay.}. 
Tab. \ref{results} shows the final results of our analysis, obtained after having refined invariant mass cuts\footnote{For $M_{G^*}=1.5(2)$TeV: $M_{all}>1.3(1.7)$TeV and at least one of the conditions $M_{Wb}>0.8 (1.1)$TeV, $M_{Wt}>0.8 (1.1)$TeV respected 
[we also refine the cut $p_{Tj(1)}>275 (400)$GeV ].\\ 
For $M_{G^*}=3 (4)$TeV: $M_{all}>2.7(3.6)$TeV and one of the conditions $1.4(2.4)$TeV $<M_{Wb}>2.6 (3)$TeV, $M_{Wt}>1.4 (2.4)$TeV respected. We also
considered in the analysis a b-tag efficiency of $60\%$ and a $1/100$ rejection factor.}. Our results
show the possibility for both the $G^*$ (with masses up to $\simeq 2$ TeV) and heavy fermions discovery at the early stage of the LHC. 
A component of the Signal (which gives a bump around the top mass in the $M_{Wb}$ distribution) comes from the $G^*\to t\bar{t}$ decay. 
Because the significance of this component depends on the top degree of compositeness
, we could extract hints on the value of this parameter.
\begin{table}
 \begin{tabular}[]{ccccc}
&\multicolumn{2}{c}{ $\sqrt{s} = 14$ TeV } &\multicolumn{2}{c}{ $\sqrt{s} = 7$ TeV }\\
\hline
& $\mathcal{L}_{5 \sigma}$ & $S/B$ &$\mathcal{L}_{5 \sigma}$ & $S/B$\\
\hline
$M_{G^*}=1.5$ TeV & 38 pb$^{-1}$ & 8.0 & 0.30 fb$^{-1}$& 9.6 \\
$M_{G^*}=2$ TeV & 188 pb$^{-1}$ & 12 & 2.7 fb$^{-1}$& 8.6 \\
$M_{G^*}=3$ TeV & 2.7 fb$^{-1}$ & 7.0 & &  \\
$M_{G^*}=4$ TeV & 42 fb$^{-1}$ & 4.6 & &  \\
\end{tabular}
\caption{Final Results. $\mathcal{L}_{5 \sigma}$ denotes the integrated luminosity needed for a $5\sigma$ discovery 
at the LHC, $S/B$ the Signal/Background ratio.
}
\label{results}
\end{table}
\acknowledgments
I thank R. Contino for collaboration in this work and C. Bini for useful suggestions and discussions. 
A more detailed analysis is in progress in collaboration with C. Bini, R. Contino and A. Tarola Parisse.

\end{document}